\documentclass[10pt]{article}
\font\smallit=cmti10

\usepackage{lmodern}
\usepackage[T1]{fontenc}
\usepackage[utf8]{inputenc}

\usepackage[export]{adjustbox}
\usepackage{forest}
\usepackage{amssymb,amsmath,amsthm,enumitem} 
\usepackage{subcaption,booktabs}
\usepackage{amstext}
\usepackage{amsfonts}
\usepackage{array}
\usepackage{hyperref}
\usepackage{url}
\usepackage{indentfirst}
\usepackage{tikz}
\usepackage{booktabs}
\usepackage{filecontents}
\usepackage[margin=1.5in]{geometry}    % For reducing margin
\usepackage{graphicx}
\usepackage[utf8]{inputenc}
\usepackage{hyperref}
\usepackage{caption}
\usetikzlibrary{shapes.multipart}
\tikzset{block/.style={
        font=\sffamily,
        draw=black,
        thin,
        fill=white!50,
        rectangle split,
        rectangle split horizontal,
        rectangle split parts=#1,
        outer sep=0pt},
        }

\usepackage[justification=centering]{caption} % Figures caption
\usepackage{caption}
\usetikzlibrary{arrows.meta}

\makeatletter

\renewcommand\section{\@startsection {section}{1}{\z@}
{-30pt \@plus -1ex \@minus -.2ex}
{2.3ex \@plus.2ex}
{\normalfont\normalsize\bfseries\boldmath}}

\renewcommand\subsection{\@startsection{subsection}{2}{\z@}
{-3.25ex\@plus -1ex \@minus -.2ex}
{1.5ex \@plus .2ex}
{\normalfont\normalsize\bfseries\boldmath}}

\renewcommand{\@seccntformat}[1]{\csname the#1\endcsname. }

\makeatother

\newtheorem{example}{Example}

\newtheorem{observation}{Observation}

\theoremstyle{definition}

\newtheorem{remark}{Remark}

\title{Critical issues with the Pearson’s chi-square test} 

\begin{document}
\maketitle

\vskip 15pt
\begin{center}
{\bf Vladimir Gurvich}\\
{\smallit National Research University Higher School of Economics 
(HSE), Moscow, Russia, 
RUTCOR, Rutgers,the State University of NJ, USA}\\
{\tt vgurvich@hse.ru}, {\tt vladimir.gurvich@gmail.com}\\
\vskip 10pt

{\bf Mariya Naumova}\\
{\smallit Rutgers Business School, Rutgers University, Piscataway, NJ, United States}\\
{\tt mnaumova@business.rutgers.edu}\\
\end{center}

\begin{abstract} 
Pearson's chi-square tests are among the most commonly applied statistical tools across a wide range of scientific disciplines, including medicine, engineering, biology, sociology, marketing and business. However, its usage in some areas is not correct.

For example, the chi-square test for homogeneity of proportions (that is, comparing proportions across groups in a contingency table) is frequently used to verify if the rows of a given nonnegative $m \times n$ (contingency) matrix $A$ are proportional. The null-hypothesis $H_0$: ``$m$ rows are proportional'' (for the whole population)  is rejected with confidence level  $1 - \alpha$ if and only if  $\chi^2_{stat} > \chi^2_{crit}$, 
where the first term is given by Pearson's formula, 
while the second one depends only on $m, n$, and $\alpha$, 
but not on the entries of $A$. 

It is immediate to notice that the Pearson's formula is not invariant. 
More precisely, whenever we multiply all entries of $A$ 
by a constant  $c$, 
the value  $\chi^2_{stat}(A)$ is  multiplied by $c$, too,  
$\chi^2_{stat}(cA) = c \chi^2_{stat} (A)$. 
Thus, if all rows of  $A$ are exactly proportional 
then $\chi^2_{stat}(cA) = c \chi^2_{stat}(A) = 0$ for any $c$  and any $\alpha$. 
Otherwise, $\chi^2_{stat} (cA)$ becomes arbitrary large or small, 
as positive $c$ is increasing or decreasing. Hence, at any fixed significance level $\alpha$, the null hypothesis 
$H_0$ will be rejected  with confidence $1 - \alpha$, 
when $c$ is sufficiently large and not rejected  
when $c$ is sufficiently small,  
Yet, obviously, the rows of $cA$ should be proportional or not for all $c$ simultaneously. 

Thus, any reasonable formula for the test statistic must be invariant, that is, 
take the same value for matrices $cA$ for all real positive $c$.


\end{abstract} 

\section{Introduction} 
The chi-square test for proportionality determines whether there is a statistically significant difference between the proportions of a certain outcome in two independent groups.

The original chi-square test, commonly referred to as Pearson’s chi-square, originated from Karl Pearson’s papers in the late 1800 - early 1900s \cite{Pea1894, Pea1895, Pea1897, Pea1900}. 
It is used both as a ``goodness of fit test'' - where data are classified along a single dimension - and as a test for contingency tables, where classification occurs across two or more dimensions. 
A historical overview of the development of the test is provided in \cite{Pla83, Coch52}. Pearson suggested but did not provide a proof that the test statistic follows the chi-square distribution \cite{Coch52}. The correction regarding the number of degrees of freedom was addressed in Fisher’s papers published in 1922 and 1924 \cite{Fish22, Fish24}.

Currently, the chi-square test is extensively utilized across a broad spectrum of research disciplines in analyzing categorical data. Its applications span fields such as social sciences, biomedical research, economics, education, and marketing, where it is employed to assess associations between variables and evaluate the goodness of fit of observed data to expected distributions. This widespread adoption underscores its substantial role in empirical research and statistical inference. According to \cite{Bai83}, the chi-square test has seen more use than any other statistical test beside possibly Student's t-test.

However, despite its widespread application, multiple authors (see for ex., \cite{Coch52, FHC12, Kos21, McH13}) mention that the chi-square test presents several methodological limitations, such as (not theoretically proven) lower and upper bounds on the sample size, the assumption that the data are derived from independent observations, etc. 

In this paper, we demonstrate an extreme sensitivity of the test to the sample sizes and data representation. Specifically, we show that if the given contingency matrix is multiplied by a constant $c$, the test statistic is multiplied by $c$ as well. The results of the test can therefore be easily manipulated (to reject or not reject $H_0$).

\begin{remark}
Similar problems were recently detected for the analysis of variance (ANOVA) and the Tukey-Kramer test.  
Tukey-Kramer's formula [see (4) in \cite{GN21}] of the critical range for groups of observations $i$  and $j$ depends not only on these two groups but also on all other groups. This contradicts common sense, since these other groups may be not related to groups $i$ and $j$ at all, and immediately leads to some logical contradictions in ANOVA with more than two groups of observations; see \cite{GN21}  for definitions and more details. 
\end{remark}

\section{Pearson Chi-Square Test for Homogeneity of Proportions}
All three omnibus tests in the Pearson family - goodness of fit, independence, and homogeneity of proportions - share basically the same underlying formula for the test statistic. We will concentrate on the homogeneity test.

Consider a scenario in which $T$ outcomes of multinomial trials are classified according to two distinct criteria, into one of the same $m$ categories, in one of $n$ groups. The outcomes can be presented in a two-way contingency table with $m$ rows and $n$ columns, as shown in Table \ref{cont_table}.
\begin{table}[h!]
\centering
\begin{tabular}{lcccccc|c}
\toprule
 & Group 1 & Group 2 & $\cdots$ & Group $j$ & $\cdots$ & Group $n$ & Row Total \\
\midrule
Category 1 & $O_{11}$ & $O_{12}$ & $\cdots$ & $O_{1j}$ & $\cdots$ & $O_{1n}$ & $O_{1\cdot}$ \\
Category 2 & $O_{21}$ & $O_{22}$ & $\cdots$ & $O_{2j}$ & $\cdots$ & $O_{2n}$ & $O_{2\cdot}$ \\
$\vdots$ & $\vdots$ & $\vdots$ & & $\vdots$ & & $\vdots$ & $\vdots$ \\
Category $i$ & $O_{i1}$ & $O_{i2}$ & $\cdots$ & $O_{ij}$ & $\cdots$ & $O_{in}$ & $O_{i\cdot}$ \\
$\vdots$ & $\vdots$ & $\vdots$ & & $\vdots$ & & $\vdots$ & $\vdots$ \\
Category $m$ & $O_{m1}$ & $O_{m2}$ & $\cdots$ & $O_{mj}$ & $\cdots$ & $O_{mn}$ & $O_{m\cdot}$ \\
\midrule
Column Total & $O_{\cdot1}$ & $O_{\cdot2}$ & $\cdots$ & $O_{\cdot j}$ & $\cdots$ & $O_{\cdot n}$ & $T$ \\
\bottomrule
\end{tabular}
\caption{General structure of an $m \times n$ contingency table. Here, $O_{ij}$ denotes the observed (absolute) frequency in cell $(i, j)$, $O_{i\cdot}$, $O_{\cdot j}$ denote the marginal totals for row $i$ and column $j$ respectively, and $T$ is the grand total.}
\label{cont_table}
\end{table}

For example, in \cite{FHHW90}, a population-based case-control study of prostatic cancer in Alberta uses the contingency table related to 376 newly diagnosed prostatic cancer patients and 620 controls, group-matched based on their ethnicity (see Table \ref{tab_cancer}).

\begin{table}[h!]
\centering
\begin{tabular}{lcccccc|c}
\toprule
 & Cases & Controls & Row Total \\
\midrule
British & $200$ & $279$ & $479$ \\
French & $16$ & $20$  & $36$\\
German & $55$ & $93$  & $148$\\
Ukrainian  & $31$ & $79$  & $110$\\
Others & $74$ & $149$  & $223$\\
\midrule
Column Total & $376$ & $620$ & $996$ \\
\bottomrule
\end{tabular}
\caption{The contingency table for a population-based case-control study of prostatic cancer in Alberta, with 376 newly diagnosed prostatic cancer patients and 620 controls of different ethnicities \cite{FHHW90}}
\label{tab_cancer}
\end{table}

Let $O_{ij}$ denote the observed (absolute) frequency in the cell corresponding to row $i$ and column $j$, for $i = 1, 2, \dots, m$ and $j = 1, 2, \dots, n$.

Let
\begin{align*}
R_i &= \sum_{j=1}^{n} O_{ij}, \hspace{2mm} i \in \{1, 2, ..., m\}& \text{(row sum for population } i\text{)},\\
C_j &= \sum_{i=1}^{m} O_{ij}, \hspace{2mm} j \in \{1, 2, ..., n\} & \text{(column sum for category } j\text{)} \\
T &= \sum_{i=1}^{m} \sum_{j=1}^{n} O_{ij}. & \text{(total sample size)}
\end{align*}

The null hypothesis is:
\[
H_0: \pi_{1j} = \pi_{2j} = \cdots = \pi_{mj}, \quad \forall j \in \{1, \dots, n\},
\]
where \( \pi_{ij} \in [0,1] \) denotes the proportion of individuals in group \( i \) falling into category \( j \), and \( \sum_{j=1}^{n} \pi_{ij} = 1 \) for each \( i \).

Under \( H_0 \), the expected frequency \( E_{ij} \) in cell \( (i,j) \) is given by:
\[
E_{ij} = \frac{R_i \cdot C_j}{N}, \quad \forall i = 1,\dots,m; \quad j = 1,\dots,n.
\]

Define the test statistic:

\begin{equation}
\label{eq-Pearson}
\chi^2_{stat} = \sum_{i=1}^{m} \sum_{j=1}^{n} \frac{(O_{ij} - E_{ij})^2}{E_{ij}}.
\end{equation}

Under the test assumptions (see Section 3), the test statistic \( \chi^2_{stat} \) approximately follows a chi-square distribution with \( (m - 1)(n - 1) \) degrees of freedom under \( H_0 \). Let \( \alpha \in (0,1) \) be the significance level. 
Then the decision rule is:
\[
\text{Reject } H_0 \quad \text{if} \quad \chi_{stat}^2 > \chi^2_{1 - \alpha, (m - 1)(n - 1)},
\]
where \( \chi^2_{1 - \alpha, (m - 1)(n - 1)} \) is the \( (1 - \alpha) \)-quantile of the chi-square distribution with \( (m - 1)(n - 1) \) degrees of freedom.

\section{Assumptions of the Chi-Square Test}

While the mathematical formulation of the chi-square test statistic is relatively simple - comparing observed frequencies with expected frequencies under a specified null hypothesis- the validity of the test's conclusions rests on a set of non-trivial assumptions. These assumptions pertain to the way the data are generated, the structure of the contingency table, and the theoretical conditions under which the chi-square distribution serves as a valid approximation to the true sampling distribution of the test statistic.

Specifically, the assumptions of the test are as follows:

\begin{enumerate}[label=(\roman*)]
    \item \textbf{Random Sampling:} The data must be drawn via random sampling.

    \item \textbf{Categorical Data:} The variables involved must be categorical. The full set of \( m \times n \) cells represents all possible combinations of category levels, and each observation must belong to exactly one cell.

    \item \textbf{Expected Frequency Condition:} The expected frequency in each cell should be sufficiently large to ensure the validity of the chi-square approximation. Specifically, it is recommended that for each cell \( (i,j) \), 
    \[
        E_{ij} \geq 5\footnote{In some sources, \( E_{ij} \geq 10 \) \cite{Coch52}, but neither requirement seems to be mathematically justified.}, \quad \forall\, i \in \{1, \dots, m\},\ j \in \{1, \dots, n\}.
    \]

    \item \textbf{Independence of Observations:} Each observation must contribute to exactly one cell, and observations must be independent. That is,
    \[
        \text{Cov}(O_{ij}, O_{kl}) = 0 \quad \text{for all } (i,j) \ne (k,l).
    \]

    \item \textbf{Fixed Margins (if applicable):} For the test of homogeneity or independence, it is often assumed that either the row totals, the column totals, or both are fixed by the sampling design or conditioning.

\end{enumerate}

\section{Only invariant tests can verify proportionality}

Obviously, $\chi^2_{stat}(cA) = c \chi^2_{stat}(A)$ for all real $c \geq 0$. 

The null hypothesis $H_0$ stated in Section 2 is rejected for any fixed significance level $\alpha$, 
if and only if  $\chi^2_{stat}(A) > \chi^2_{crit}(A)$, where $\chi^2_{crit}(A) = \chi^2_{1 - \alpha, (m - 1)(n - 1)}$ is the \( (1 - \alpha) \)-quantile of the chi-square distribution with \( (m - 1)(n - 1) \) degrees of freedom, which depends only on $m, n$, and $\alpha$, but not on the entries of $A$. 

\begin{observation}
The following five statements are equivalent: 
\begin{enumerate}[label=(\alph*)]

\item the rows of $A$  are exactly proportional; 

\item the columns of $A$ are exactly proportional; 

\item  $O_{ij} = E_{ij}$  for all  
$i = 1, \dots, m$ and $j = 1, \dots, n$;  

\item  $\chi^2_{stat}(A) = 0$;  

\item  $\chi^2_{stat}(cA) = 0$ for all real positive $c$.      
\end{enumerate}
\end{observation}

Furthermore, if statements (\textit{a - e}) fail, then $\chi^2_{stat}(A) > 0$ and 
for any fixed significance level $\alpha$, the null hypothesis $H_0$  will be rejected 
for  $cA$  with confidence $1 - \alpha$, if $c$ is sufficiently large, and it will not be rejected  if $c$ is sufficiently small.  

\begin{proof}
It is enough to notice that in the numerator and denominator of (\ref{eq-Pearson}) are, respectively, a quadratic and linear functions of the entries of $A$. 
Hence, $\chi^2_{stat}(cA) = c\chi^2_{stat}(A)$.     
\end{proof} 

Yet, obviously, the result of testing proportionality of the rows of $cA$ should not linearly depend on $c$. 

\section{Examples} 
\label{examples} 

\begin{example}
We consider the following $2 \times 2$ contingency table:

\[
\begin{array}{c|cc|c}
 & \text{Group 1} & \text{Group 2} & \text{Row Total} \\
\hline
\text{Category A} & 1 & 1 & 2 \\
\text{Category B} & 1 & 11 & 12 \\
\hline
\text{Column Total} & 2 & 12 & 14 \\
\end{array}
\]

We test the null hypothesis that the proportions of outcomes are the same across the two groups.

The expected value for each cell under the null hypothesis are
\[
E_{11} = \frac{2 \cdot 2}{14} = \frac{4}{14} = \frac{2}{7}, \quad
E_{12} = \frac{2 \cdot 12}{14} = \frac{24}{14} = \frac{12}{7},
\]
\[
E_{21} = \frac{12 \cdot 2}{14} = \frac{24}{14} = \frac{12}{7}, \quad
E_{22} = \frac{12 \cdot 12}{14} = \frac{144}{14} = \frac{72}{7},
\]

and the chi-square statistic is found to be
\[
\chi^2_{stat} = \frac{(1 - \frac{2}{7})^2}{\frac{2}{7}} + \frac{(1 - \frac{12}{7})^2}{\frac{12}{7}} + \frac{(1 - \frac{12}{7})^2}{\frac{12}{7}} + \frac{(11 - \frac{72}{7})^2}{\frac{72}{7}} = \frac{25}{7} \left(\frac{1}{2}+\frac{1}{6} + \frac{1}{72}\right) = \frac{175}{72}.
\]

With $1$ degree of freedom, and using the critical value from the Chi-square distribution table at significance level $\alpha = 0.05$, which is approximately $3.841$, we have that

\[
\chi^2_{stat} \approx 2.43 < 3.841
\]

Thus, we fail to reject the null hypothesis. There is not enough evidence to suggest a significant difference in proportions between the two groups.

Let's modify the original table by multiplying each entry by $2$:

\[
\begin{array}{c|cc|c}
 & \text{Group 1} & \text{Group 2} & \text{Row Total} \\
\hline
\text{Category A} & 2 & 2 & 4 \\
\text{Category B} & 2 & 22 & 24 \\
\hline
\text{Column Total} & 4 & 24 & 28 \\
\end{array}
\]

Performing the same operations, we find the expected counts under the null hypothesis:
\[
E_{11} = \frac{4 \cdot 4}{28} = \frac{16}{28} = \frac{4}{7}, \quad
E_{12} = \frac{4 \cdot 24}{28} = \frac{96}{28} = \frac{24}{7},
\]
\[
E_{21} = \frac{24 \cdot 4}{28} = \frac{96}{28} = \frac{24}{7}, \quad
E_{22} = \frac{24 \cdot 24}{28} = \frac{576}{28} = \frac{144}{7}.
\]

Then the chi-square statistic is computed as

\[
\chi^2_{stat} = \frac{(2 - \frac{4}{7})^2}{\frac{4}{7}} + \frac{(2 - \frac{24}{7})^2}{\frac{24}{7}} + \frac{(2 - \frac{24}{7})^2}{\frac{24}{7}} + \frac{(22 - \frac{144}{7})^2}{\frac{144}{7}} = \frac{100}{7}\left(\frac{1}{4}+\frac{1}{12}+\frac{1}{144}\right) = \frac{175}{36}.
\]

Comparing this value to the critical value \( \chi^2_{0.05,1} = 3.841 \), we see that \[
\chi^2_{stat} \approx 4.86 > 3.841,
\]
and reject the null hypothesis.

There is sufficient evidence to suggest that the proportions differ between the two groups.
\end{example}

Vectors $(1, 1)$ and $(1, 11)$ do not look proportional at all. 
So it seems a bit strange 
that we cannot reject proportionality, for the population,  
with confidence 95\%, based on such a sample. 

Some researchers (see, for ex. \cite{Coch52, GN96, FHC12}] recommend to apply the Pearson's test only to samples with ``large enough'' entries. In particular, the expected frequencies in each cell should generally be at least ten, or in some sources, no less than 5 (both numbers seem to be arbitrarily chosen) \cite{Coch52}, or otherwise, the test can result in inaccurate p-values and reduced statistical power. Indeed, doubling the values in the first sample, we obtain the second one, with row vectors (2, 2) and (2, 22), which already enables us to reject proportionality with confidence 95\%. 

However, one should note that with very large samples, even trivial associations may yield statistically significant results, potentially leading to overinterpretation. 

These restrictions are ambiguous and do not solve the problem in general, since, as we already mentioned, $\chi^2_{stat}(cA) \equiv c \chi^2_{stat}(A)$. 

\begin{example}
We illustrate the test for the following $2 \times 2$ contingency table:

\[
\begin{array}{ccc|c}
 & \text{Group A} & \text{Group B} & \text{Row Total} \\
\hline
\text{Category 1} & 22 & 18 & 40 \\
\text{Category 2} & 18 & 22 & 40 \\
\hline
\text{Column Total} & 40 & 40 & 80 \\
\end{array}
\]

Under the null hypothesis of homogeneity, expected counts for all cells are
\[
E_{ij} = \frac{40 \cdot 40}{80} = 20, i, j \in \{1, 2\}.
\]

The chi-square statistic:
\[
\chi^2_{stat} = \frac{(22 - 20)^2}{20} + \frac{(18 - 20)^2}{20} + \frac{(18 - 20)^2}{20} + \frac{(22 - 20)^2}{20} = \frac{4}{5} = 0.8.
\]

Since \(0.8 < 3.841 \), we do not reject the null hypothesis. 
There is no significant difference in proportions between the groups.

If we modify the original contingency table by multiplying all elements by 1,000 (for example, assuming that the elements of the matrix represent the actual number of cases while before they represented the number of cases in thousands). This yields the following scaled observed frequencies:

\[
\begin{array}{ccc|c}
 & \text{Group A} & \text{Group B} & \text{Row Total} \\
\hline
\text{Category 1} & 22000 & 18000 & 40000 \\
\text{Category 2} & 18000 & 22000 & 40000 \\
\hline
\text{Column Total} & 40000 & 40000 & 80000 \\
\end{array}
\]

Under the null hypothesis of homogeneity, all expected counts are
\[
E_{ij} = \frac{40000 \cdot 40000}{80000} = 20000, i, j \in \{1, 2\}, 
\]

and the chi-square statistic:
\[
\chi^2_{stat} = \frac{(22000 - 20000)^2}{20000} + \frac{(18000 - 20000)^2}{20000} + \frac{(18000 - 20000)^2}{20000} + \frac{(22000 - 20000)^2}{20000} = 800.
\]

Note that the new $\chi^2_{stat}$ is $1,000$ times greater than the original one.

Comparing $\chi^2_{stat}$  to the critical value ($800 > 3.841$), we reject the null hypothesis. There is a significant difference in proportions between the groups.

\end{example}

\begin{example}
In this case, we work with the following $3 \times 4$ contingency matrix of observed frequencies:

\[
\begin{array}{c|cccc|c}
 & \text{Group 1} & \text{Group 2}& \text{Group 3}& \text{Group 4} & \text{Row Total} \\
\hline
\text{Category A} & 98 & 86 & 79 & 71& 334\\
\text{Category B} & 78 & 82 & 88 &51 &299\\
\text{Category C} & 75 & 62 & 82 & 77& 296\\
\hline
\text{Column Total} & 251 & 230 & 249 &199 & 929\\
\end{array}
\]

The matrix of the expected frequencies is given by 

\[
E = \begin{bmatrix}
90.3 & 82.7 & 89.5 & 71.6 \\
80.9 & 73.9 & 80.2 & 64.1 \\
79.9 & 73.3 & 79.2 & 63.4
\end{bmatrix},
\]
and the chi-square statistic is found to be $\chi^2_{stat} \approx 11.475$.

For the given $3 \times 4$ contingency matrix, the number of degrees of freedom is $(3 - 1)(4 - 1) = 6.$ So, the p-value is about $0.07$ and for $\alpha = 0.05$ we don't reject the null hypothesis. In this case, the critical value is $12.59$.

It is easy to verify that multiplying all entries, 
for example, by $2$ yields the doubled test statistic 
$(\chi^2_{stat} \approx 22.95)$ and the much lower p-value $(\approx 0.0008)$.
\end{example}

\begin{example} 
\label{ex-4}
We now provide a construction of a matrix for which formula (\ref{eq-Pearson}) can be applied.
Suppose we want to analyze joint probabilistic distribution of two random variables taking  $m$ and $n$  distinct values, say, $X = (x_1, \dots, x_m)$  and  $Y = (y_1, \dots, y_n)$. Our goal is to estimate the corresponding two probabilistic distributions $(p_1, \dots, p_m), (q_1, \dots, q_n)$ and decide if $x$ and $y$  are independent. 
A given sample consists of  $T$ trials, each of which gives us the values of  $x$ and $y$, simultaneously. 
Let $Z_{ij}$  be the number of trials realizing  $x_i$  and $y_j$. 
Then, $z_{ij} = Z_{ij}/T$ are the corresponding observed frequencies. Set 
$p_i = \sum_{j=1}^n z_{ij}, \;  q_j = \sum_{i=1}^m z_{ij}$,   
and treat them as the observed frequencies for  $x$ and $y$, respectively. 
By the above definitions, we have 
$$\sum_{i=1}^m p_i = \sum_{j=1}^n q_j = \sum_{i=1}^m \sum_{j=1}^n z_{ij} = 1.$$
We define the expected frequencies  
setting  $z^e_{ij} = p_i q_j$, and apply equation (\ref{eq-Pearson}) to obtain the test statistic. 
\end{example}

Although the last example is also a case of verifying proportionality, it is special. Since the sum of all entries equals 1, multiplication of all observed frequencies $z_{ij}$  by a constant  $c \neq 1$  is impossible. 
More generally, the test will work when the sum of all entries must be constant, or almost constant. 

\section{Conclusions} 
The Pearson chi-square test is one of the most widely used tools for analyzing categorical data, particularly in contingency tables. As demonstrated through both theoretical formulation and numerical examples, even under its common assumptions, including sufficiently large expected cell frequencies and the independence of observations, the test is extremely sensitive to the input data and its inaccurate application can lead to misleading conclusions. 

Pearson's chi-square test cannot be applied for verifying proportionality for general contingency matrices. Yet, it is applicable in cases when the total sum of entries is a constant, or almost constant, for example, for testing the independence of two probabilistic distributions, given a sample of pairwise frequencies.

One could modify formula (\ref{eq-Pearson}) making it invariant with respect to multiplication of $A$ by a constant (for example, by squaring all denominators in (\ref{eq-Pearson}) or by dividing all entries of $A$ by their sum or their maximum). However, in this case, the distribution $\chi^2_{crit}$ should be modified accordingly. 
Both modifications deserve a separate consideration.

\section*{Acknowledgements} 
This research was supported by the Russian Science Foundation, grant 20-11-20203.

\end{document}